# SL5 Standard for AI Security

Preliminary Draft

Version 0.1

Lisa Thiergart

Yoav Tzfati

Peter Wagstaff

Guy

Luis Cosio

Philip Reiner

# Table of Contents

# About the Security Level 5 Task Force

The SL5 Task Force is a non-profit cross-industry effort working to ensure frontier AI infrastructure can achieve nation-state-level security by 2028/2029. Founded in March 2025, we are a core team of engineers and security strategists leading a 100-person technical track (comprising security engineers from frontier AI labs, government security specialists, and datacenter colocation providers) alongside an executive track of AI industry security leaders providing steering input.

Over the past nine months, we have conducted a series of workshops and research programs to clarify what it takes to reach Security Level 5 in a way that is sensitive to competitive pressures, the need to maintain speed of innovation, and the reality of a rapidly shifting threat landscape. Our mission is to create the optionality for frontier AI labs to reach Security Level 5 in the coming years, and to be able to activate that security level within six months of choosing to do so.

In service of that mission, we have convened this broad task force to clarify what needs to be done, and in particular what needs to be done early, to preserve that optionality. This standard represents one output of that collaborative effort.

For more information or to engage with our work, contact us at [standard@sl5.org](mailto:standard@sl5.org) or visit [sl5.org](https://sl5.org).

# About This Document

Security Level 5 (SL5) is a security posture for AI systems that could plausibly thwart top-priority operations by the world's most cyber-capable institutions: those with extensive resources, state-level infrastructure, and expertise years ahead of the public state of the art. The SL5 terminology originates from the RAND Corporation's 2024 report "Securing AI Model Weights" [1].

This first revision of the SL5 standard focuses on requirements with long lead times: interventions that must be planned years in advance, such as facility construction, hardware procurement, and organizational capability development. We prioritize these requirements because preserving optionality for SL5 by 2028/2029 requires starting now. These capabilities cannot be retrofitted on short notice when the need becomes urgent. Some requirements represent significant departures from current day standard practice. We believe bold measures are necessary for this level of security and see clear opportunities to apply optimization pressure to existing and novel solutions to customize them for the AI industry and address the practical operational requirements as much as possible. Our organization exists to begin paving this path. Some requirements approximate government security capabilities where private-sector approaches may be insufficient. We identify these gaps and note where government involvement may ultimately be necessary.

This standard was developed collaboratively with frontier AI laboratories, government partners, and security experts through sustained engagement over several months. As version 0.1, significant refinement is expected through continued stakeholder engagement. We explicitly invite frontier AI labs, government agencies, datacenter operators, and security researchers to engage with this work, whether through direct collaboration, feedback, or implementation experience. Please reach out through standard@sl5.org.

The control specifications in this standard are structured as an overlay on NIST SP 800-53. We chose this approach for three reasons. First, NIST SP 800-53 is a battle-tested framework and the standard choice for high-security organizations. Second, structuring as an overlay enables ease of adoption for organizations already implementing NIST controls. Third, the overlay format clearly expresses the "diff" from existing baselines, highlighting what is new or different for SL5 rather than restating established requirements.

Many other security controls are necessary for SL5, including most controls from existing high-security baselines. Future revisions will provide detailed mapping from DoD Impact Level 6 (IL6) and its reference frameworks (FedRAMP High, CNSSI 1253) to SL5 requirements [4], [5]. Physical security requirements draw on ICD 705 SCIF standards as a basis [6], [7], [22], [23]. Hardware supply chain requirements reference NIST SP 800-161 Rev 1 [3].

# Acknowledgments

This standard was shaped by the collective expertise of the SL5 Task Force's two specialized tracks. Our Executive Track, comprising AI lab decision-makers, national security leaders, datacenter operators, chip providers, and government representatives, provided strategic direction and identified key obstacles to SL5 implementation. Our Technical Track, comprising security researchers, lab security engineering staff, technical AI researchers, hardware engineers, and security engineers working across five specialized working groups (machine security, network security, software security, supply chain security, and personnel security), developed and stress-tested the control specifications and architectural recommendations in this document.

We are deeply grateful to every participant across both tracks for their willingness to engage in rigorous, honest discussion and for the time they generously contributed to this effort. Please note that the following list of acknowledgments is not yet finalized and will be updated in the coming days. Additionally, inclusion in this list reflects an individual's informative contributions to this effort and does not necessarily represent their personal views or endorsement of all the contents of this document. The following individuals wished to be acknowledged for their contributions:

- Jason Clinton, Deputy CISO, Anthropic
- Dane Stuckey, CISO, OpenAI
- Vijay Bolina, former CISO, GDM
- Rob Joyce, former Cybersecurity Director, NSA
- Phil Venables, former CISO, Google Cloud, current Partner at Ballistic Ventures
- Jason Kichen, CISO, Fluidstack
- Eric Grosse, former VP Security & Privacy Engineering, Google
- Dominic Rizzo, CEO ZeroRISC
- Omer Nevo, CTO, Irregular
- Evan Miyazono, CEO, Atlas Computing
- Kristian Rönn, CEO, Lucid Computing
- Jarrah Bloomfield, Anthropic & former Google
- Keri Warr, Anthropic
- Steven Hernandez, former CISO, United States Agency for International Development
- Tanya Verma, CEO, Tinfoil
- Gil Gekker, CoS to CEO, Irregular
- davidad (David A. Dalrymple)
- Paul Crowley (ciphergoth), Anthropic
- Jason Kikta, CTO, Automox

*This list will be in progress for the next few weeks, as we hear from members.*

# 1. Introduction

## 1.1 Threat Model

The primary adversaries for SL5 systems are the top-priority operations of the world's most cyber-capable institutions—operations comparable to 1,000 individuals with expertise years ahead of the public state of the art, spending years with budgets up to $1 billion, backed by state-level infrastructure and access developed over decades. A **primary use case for SL5 is an AI lab approaching fully automated AI R&D**. OpenAI has already announced they expect to have automated researchers by 2028 [29]. Anthropic's CEO has similarly noted that AI "may be only 1–2 years away from a point where the current generation of AI autonomously builds the next" [30]. At this stage, the economic value of frontier AI models derives increasingly from conducting internal automated R&D and accelerating scientific advances, more so than from serving the model to customers. Models capable of automating AI research could enable rapid recursive improvement in AI capabilities, with huge economic potential and significant implications for geopolitics. Nation-state adversaries have strong incentives to acquire such capabilities or prevent rivals from doing so.

The targets of SL5 protection are critical assets held by the frontier AI labs. These critical assets include "covered models"—frontier AI models passing capability thresholds designated by the organization as requiring SL5 protection, AI research and software that could enable adversaries to develop comparable capabilities, inputs (which could be used to poison or backdoor) and outputs (which could be used to distill or reverse engineer). The security objectives are the **confidentiality, integrity, and availability** of these critical assets.

The standard also addresses risks from misaligned or compromised AI models that may attempt to sabotage research or exfiltrate themselves, which constitute a distinct form of insider threat. Mitigations for this threat class substantially overlap with nation-state defenses; this revision does not include mitigations specific to misaligned AI, though future revisions will likely address this threat class more directly.

Attack vectors include insider threats, supply chain compromise, physical intrusion, network exploitation, side-channel attacks, and adversarial inputs designed to compromise AI systems.

## 1.2 Security Architecture Overview

This section summarizes the SL5 security architecture across five security streams: network, physical, machine, personnel, and supply chain.

### Network Security

The SL5 Network is air-gapped from external networks and supports model development, training, and internal deployment operations. The SL5 Network may span multiple geographically distributed facilities, including remote office locations (such as in San Francisco) and other datacenters connected via encrypted inter-facility links. All facilities must comply with all SL5 requirements including applicable ICD 705 guidance for physical security [6], [7], [22], [23].

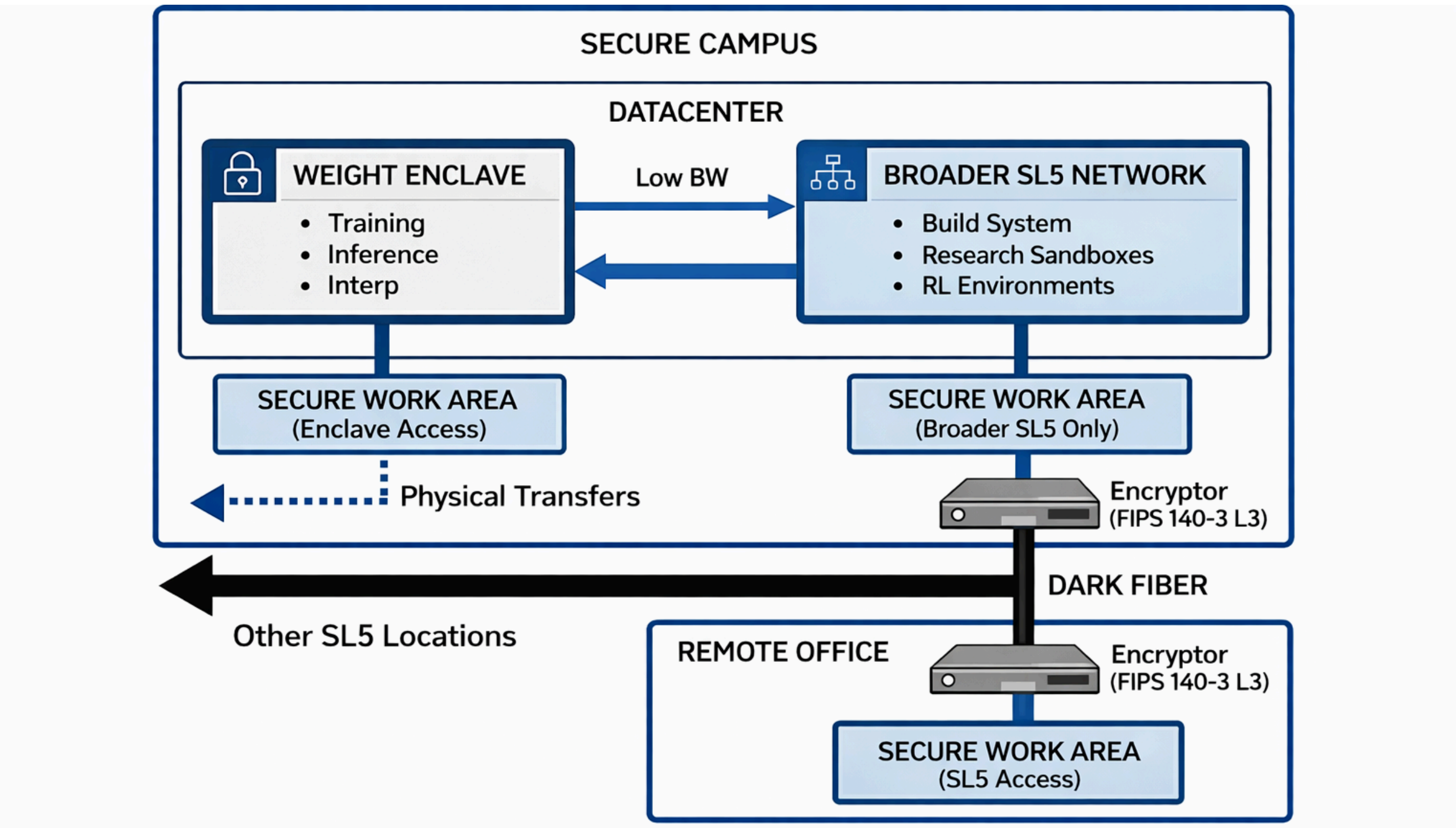

**Fig. 1.** *SL5 Network Architecture showing the relationship between Weight Enclaves, the broader SL5 Network, and inter-facility connections.*

Weight Enclaves provide additional isolation within the SL5 Network for systems with direct access to covered models. This separation enables stricter controls on the most sensitive assets—covered model weights—while supporting R&D operations and remote work locations in the broader SL5 Network. Systems requiring direct weight access (training, inference, fine-tuning, mechanistic interpretability) operate within Weight Enclaves under strict allow-by-exception code execution; all code must complete testing and signing before deployment. Research experiments conducted by AI models are executed in confined environments in the broader SL5 Network (see Figure 1), outside the Weight Enclaves. Each Weight Enclave resides in a single physical facility. Transfers exceeding the Weight Enclave outbound bandwidth limit must occur via encrypted physical media. This has significant implications for geographically distributed training: such training may prove infeasible under this constraint, or may require substantially different approaches than current practice. One speculative example is distributed RL training with daily gradient synchronization via physical media transfer. Section 1.3 highlights this as a key open question.

Key interventions:

- Air-gapped SL5 Network with no external network connections
- Weight Enclaves isolated within SL5 Network, containing only minimal necessary software with strict allow-by-exception code execution
- Single-facility restriction for Weight Enclaves; inter-enclave transfers exceeding bandwidth limits require encrypted physical media
- Dual inline network encryptors from different suppliers (per NSA "Rule of Two") for inter-facility SL5 Network connections [14]
- FIPS 140-3 Level 3 minimum validation for network encryptors [11]
- Physical bandwidth limitation on Weight Enclave boundaries preventing weight exfiltration even if other security measures fail

See Section 3.9 for detailed control specifications.

### Physical Security

SL5 facilities are constructed to an applicable subset of ICD 705 directive providing physical access control, emanations protection, and transmission security [6], [7], [22], [23]. The particular subset is determined by whether the threat model being prioritized is theft of model weights & cryptographic keys vs. also sabotage, autonomy threats and algorithmic IP theft. SL5-level security may generally warrant an updated version of ICD 705 plus potential additional AI-datacenter and AI-SCIF specific overlays. Following SCIF terminology, Red Zones are areas where SL5-protected information may be processed or stored; Black Zones are areas with no network path to SL5-protected information. All SL5 Network hardware and access locations reside within Red Zones.

Key interventions:

- ICD 705 facility construction with applicable TEMPEST countermeasures (RF shielding, power conditioning) [6], [7], [9], [23]
- Access control vestibules (mantraps) at all entry points to prevent tailgating
- Protected Distribution Systems (PDS) per CNSSI 7003 for nearby connections [8]
- Shielded rack enclosures meeting NSA 94-106 specifications for Weight Enclave systems processing covered models [10]

See Section 2 for detailed facility requirements and Section 3.4 for related control specifications.

### Machine Security

AI accelerators require hardware security features that enable protection of covered models even when the host system is compromised, when data traversing physical interconnects is intercepted, or when the hardware itself is physically attacked. Only authorized, cryptographically signed code executes on Weight Enclave accelerators (see SI-7(15)), reducing the attack surface. These capabilities enable end-to-end encrypted data paths where data arrives encrypted from origin, is decrypted only within the accelerator, and is re-encrypted before any host-accessible export.

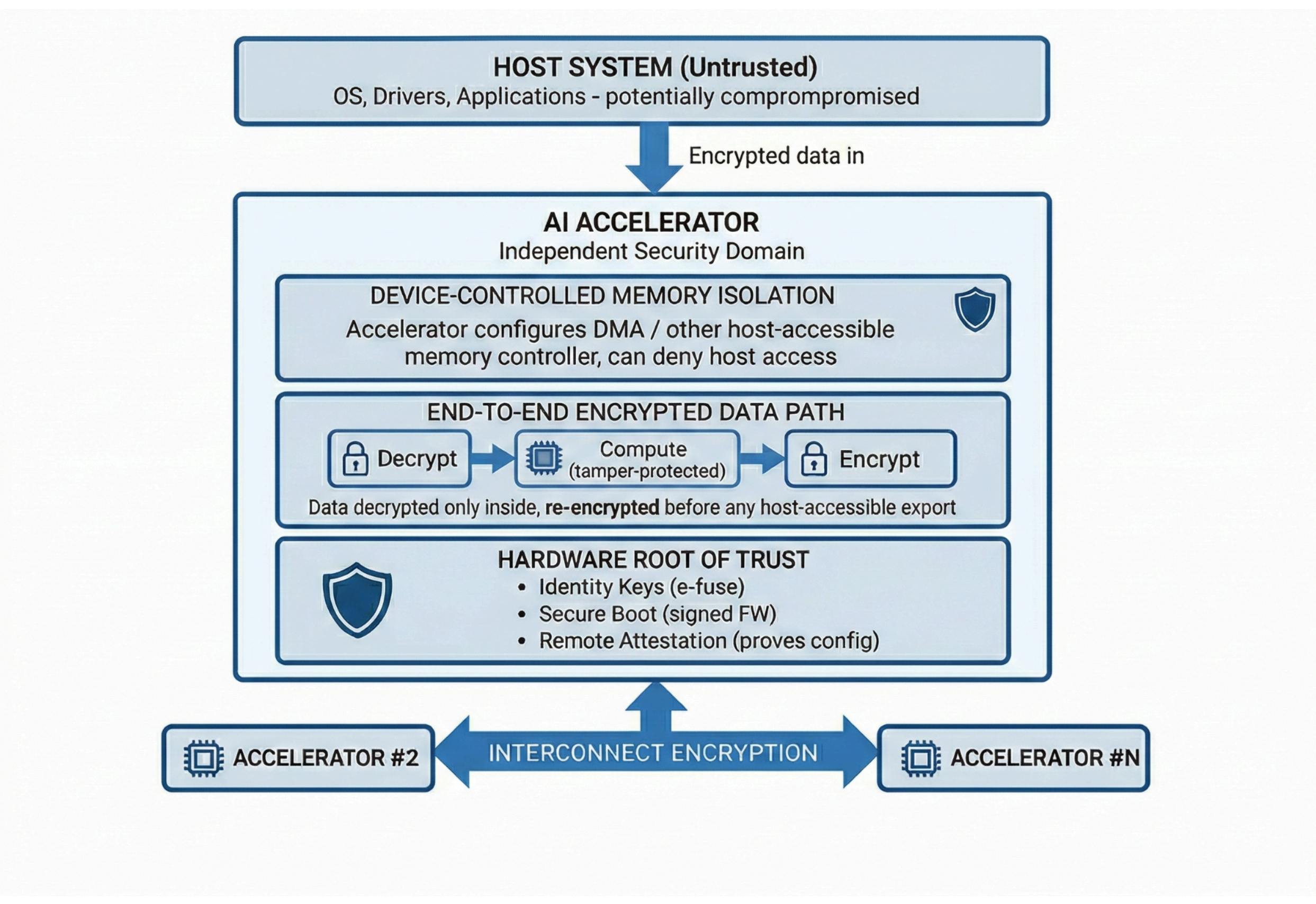

**Fig. 2.** *AI Accelerator Security Architecture showing the accelerator as an independent security domain with device-controlled memory isolation and end-to-end encrypted data paths.*

Key interventions:

- Integrated root-of-trust with hardware-provisioned identity, secure key storage, and attestation capability
- Device-controlled memory isolation allowing accelerators to deny host access independent of host software
- Interconnect encryption protecting data transmitted between accelerators during distributed operation
- Tamper protection for compute cores and accelerator memory where data exists unencrypted during computation
- Execution integrity verification ensuring only authorized code executes on accelerators

Current commercial AI accelerators vary in confidential computing capabilities, and gaps remain relative to SL5 requirements. For example, NVIDIA Blackwell's memory isolation and code verification depend on a CPU TEE rather than being accelerator-controlled, and commercial threat models exclude physical attacks [16], [24].

Current implementations also lack mechanisms preventing composition attacks, where individually-authorized operations are combined to exfiltrate data. Two approaches show promise. The Ascend-CC research architecture (2024) demonstrates confidential computing that excludes the CPU from the trusted computing base, using device-controlled memory isolation and firmware-level task sequence attestation [17]. Alternatively, restricting workloads to fused kernels verified to not exfiltrate data individually or through composition avoids the need for vendor engagement but may increase implementation complexity for organizations.

Organizations must engage accelerator vendors during architecture phases—hardware security features require years to develop, and influence over roadmaps cannot be gained retroactively.

See Sections 3.3 and 3.10 for related control specifications.

## Personnel Security

A five-tier sensitivity level framework (SenL-1 through SenL-5) addresses insider threat through graduated vetting, monitoring, and access controls. This represents the strongest feasible private-sector approximation of government clearance programs. Detailed tier definitions, vetting procedures, and operational safeguards are specified in a separate SenL Framework Document [26].

Key interventions:

- Five sensitivity levels based on access to covered models and critical infrastructure [26]
- Vetting depth scales from baseline checks (SenL-1) through verified subject and reference interviews (SenL-4/5) [26]
- Monitoring intensity scales from periodic checks to intensive monitoring of compartmented system interactions [26]
- Access obligations scale from standard NDA to custodian-specific protocols and post-employment restrictions [26]
- Dual authorization for critical operations involving covered models [26]
- SenL-5 clearance required for unescorted Red Zone access, with two-person integrity [26]

Organizations may need to pursue government involvement (through classified contract pathways, information sharing arrangements, or new statutory authority) to achieve personnel security sufficient for the stated threat model. Section 1.3 highlights this as a key open question.

See Sections 3.5 and 3.6 for detailed control specifications.

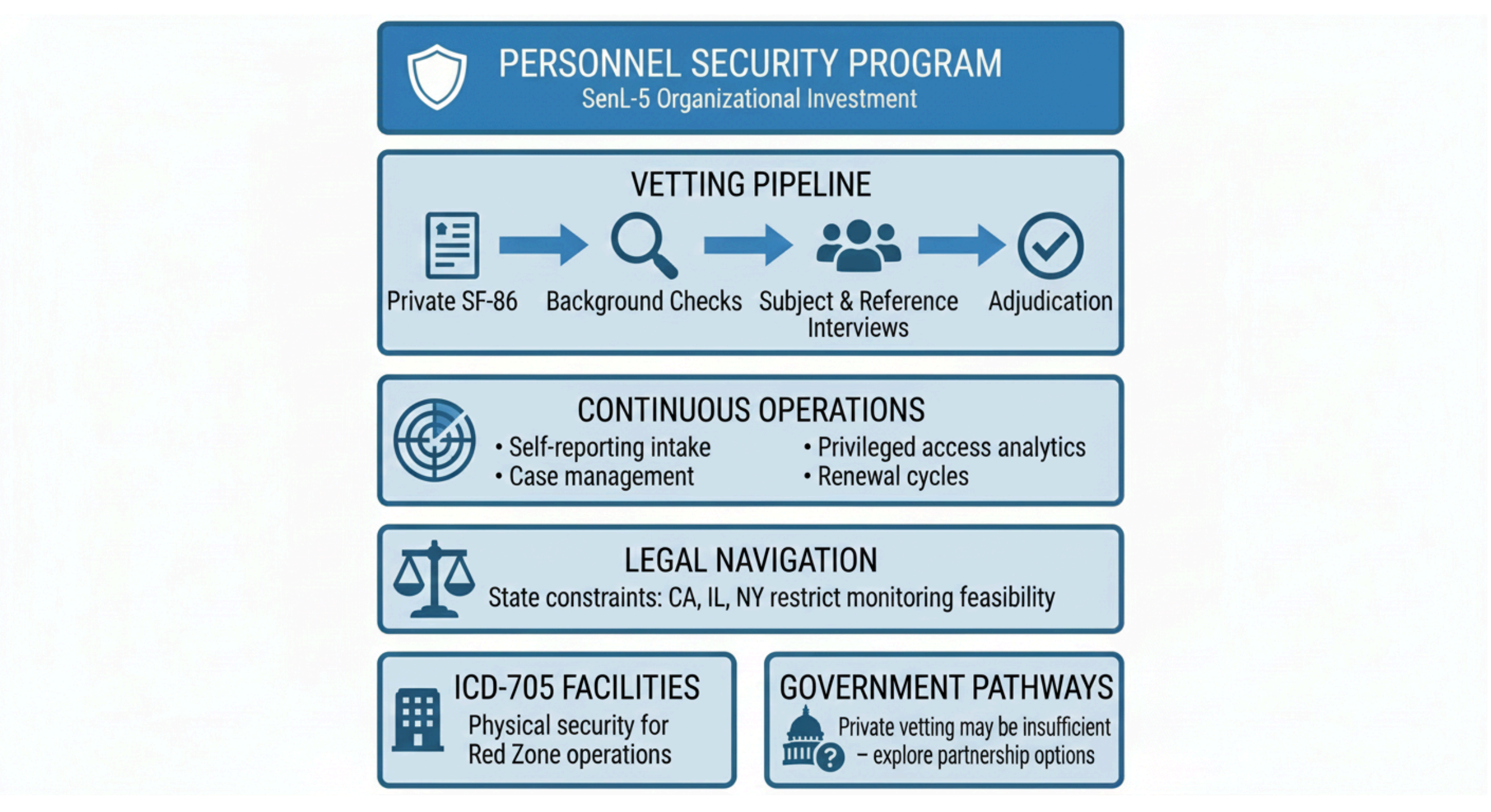

**Fig. 3.** *Personnel Security Program showing the organizational investment required for SenL-5 capability: vetting pipeline, continuous operations, legal navigation, ICD 705 facilities, and potential government pathways [26].*

## Supply Chain Security

Hardware supply chain integrity requirements address risks from compromised components. Hardware supply chain requirements follow NIST SP 800-161 Rev 1 guidance directly [3].

Data entering the SL5 environment from external sources must be screened for adversarial content—poisoning attacks, jailbreak attempts, and adversarial examples that could compromise AI systems or sabotage research. External data is a critical input to AI system development, analogous to hardware components; both can carry embedded attacks from upstream sources, and correlated attacks targeting both software and training data could amplify impact.

Robust adversarial content detection remains an open research problem; best-available defensive measures are insufficient against sophisticated adversaries [18], [19]. This standard mandates staging isolation and investment in detection research, understanding that breakthroughs in adversarial robustness are necessary to fully address this threat. Section 1.3 highlights this as a key open question.

Key interventions:

- Adversarial content screening for all external data, with staging isolation until data completes automated detection and human review
- Supplier governance: criticality-based inventory, continuous assessment against security baseline, qualified bidders/manufacturers lists, supply base diversity where feasible, requirements flow-down to sub-tier contractors
- Component integrity assurance: comprehensive testing including counterfeit detection, physical inspection, tamper detection, developer screening

See Sections 3.1 and 3.8 for detailed control specifications.

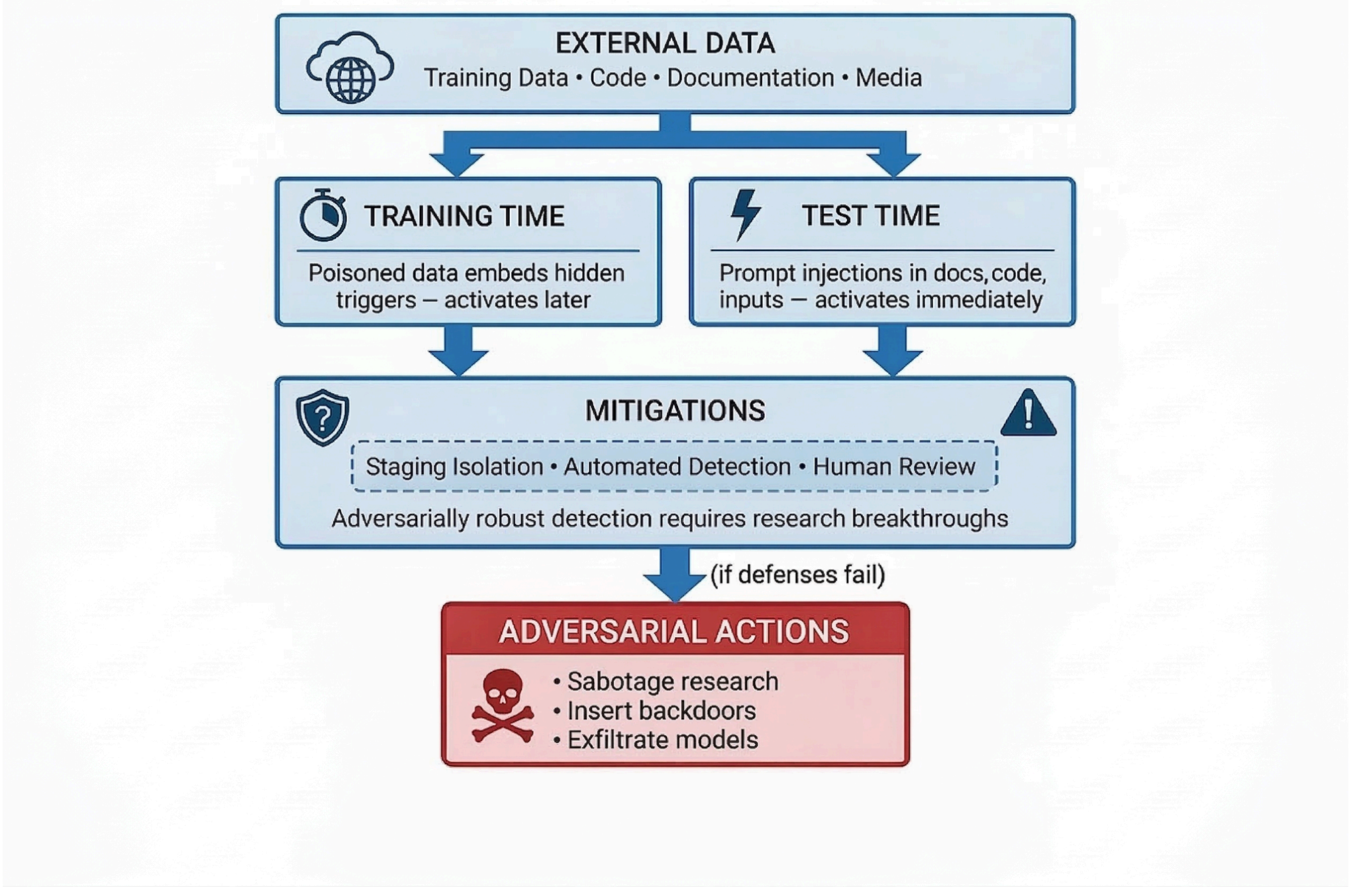

**Fig. 4.** *Adversarial Robustness threat model showing how external data can carry attacks that activate at training time or test time, the mitigations required (staging isolation, automated detection, human review), and the consequence if defenses fail.*

## 1.3 Open Questions

This is a first draft and we value transparency about areas of genuine uncertainty, whether due to conflicting expert perspectives or limited access to relevant information. Our goal is to achieve both strong security and operational effectiveness. We seek input to resolve these uncertainties directly—evidence that changes our assessment may change the requirements—as well as creative approaches that accomplish both. The following questions have significant architectural or policy implications:

1. **Personnel vetting limitations:** Whether private-sector vetting is sufficient for SL5, or whether government involvement (through classified contract pathways, information sharing, or new legal authority) is required to achieve adequate personnel security. We welcome input from organizations with experience in either approach.

2. **Adversarial detection feasibility:** Whether robust detection of adversarial content is achievable against sophisticated adversaries, given that this remains an active research problem. This standard mandates staging isolation and investment in detection research, but it is uncertain whether breakthroughs sufficient to address the threat will materialize. We welcome detection approaches or architectural mitigations.

3. **Inter-enclave network security:** Whether long-distance network connections can be adequately secured against nation-state adversaries, even with multiple validated inline encryptors in series. If such connections cannot be secured, the physical media requirement for high-bandwidth transfers would significantly constrain or preclude geographically distributed training. We seek both security analysis on network connection viability and, if the constraint stands, creative approaches to distributed training under physical media requirements.

Additional open questions are documented in Appendix A.

# 2. ICD 705 Facility Requirements

Physical security forms the foundation for all other SL5 protections [6], [7], [22], [23]. This section specifies facility construction requirements that have no direct NIST SP 800-53 equivalent. These requirements draw on an applicable subset of ICD 705 and ICS 705-1 (Physical and Technical Standards for Sensitive Compartmented Information Facilities), published by the Office of the Director of National Intelligence [6], [7], [22], [23]. The particular subset is determined by whether the threat model being prioritized is theft of model weights and cryptographic keys, or also extends to sabotage, autonomy threats, and algorithmic IP theft. SL5-level security may generally warrant an updated version of ICD 705 plus potential additional AI-datacenter and AI-SCIF specific overlays. ICD 705 provides detailed specifications for perimeter construction, intrusion detection, access control, acoustic protection, and TEMPEST countermeasures; this section summarizes key requirements and SL5-specific adaptations.

**Adaptation for SL5:**

- “SL5-protected information” (covered models and secrets) replaces “Classified Information” [26]
- Chief Security Officer (CSO) or Site Security Manager serves as accreditation authority
- SenL-5 clearance replaces government security clearances [26]

### Security in Depth

Facility design implements Security in Depth (SID)—layered security controls that increase the probability of detecting unauthorized access attempts before reaching the Red Zone perimeter. Layers may include perimeter fencing, building access controls, and controlled areas surrounding Red Zones. SID considerations should inform site selection and building layout.

### Zone Architecture

**Red Zone:** The SL5 Network operates within Red Zones—physically hardened environments constructed to ICD 705/ICS 705-1 standards. Any system with network access to SL5-protected information must be within a Red Zone. Red Zones may span multiple geographically distributed facilities.

**Black Zone:** Areas with no network path to SL5-protected information. Black Zones may exist within the same facility as Red Zones but are completely network-isolated.

### Construction

Red Zones form complete physical enclosures per ICS 705-1—walls, floor, and ceiling create a continuous barrier with no gaps. Acoustic attenuation (Sound Group 3 or 4) prevents eavesdropping through the perimeter [7], [23]. Windows are prohibited; where required by safety codes, they must be non-opening with RF/optical shielding. All utility penetrations (HVAC, power, fire suppression) must maintain perimeter integrity.

### TEMPEST

Red Zones require TEMPEST countermeasures per NSTISSAM TEMPEST/1-92, pre-engineered into construction to the maximum extent practicable [9], [21]. RF shielding prevents electromagnetic signal egress from the perimeter, dedicated power conditioning prevents power-line analysis, and all equipment must be hardwired—no wireless devices permitted.

### Intrusion Detection

Red Zones require intrusion detection systems (IDS) per ICD 705 Chapter 7, with sensors covering all perimeter entry points and motion detection within the space. Alarm response time requirements depend on storage type: 5 minutes for open storage (Sensitive compartmented information (SCI) material accessible outside containers), 15 minutes for closed storage (SCI material secured in General Services Administration (GSA)-approved containers when unoccupied) [6], [7], [23].

### Transmission Security

All Weight Enclave traffic leaving the Red Zone perimeter requires Protected Distribution Systems (PDS) per CNSSI 7003—including encrypted inter-building connections [8]. PDS provides physical protection through hardened conduit and alarmed carriers.

# 3. Control Specifications

This section is a NIST SP 800-53 overlay—a set of supplemental guidance and parameter values that tailor existing controls without replacing base requirements. It is a partial overlay, covering only the long lead time interventions highlighted in Section 1. The complete SL5 overlay will build on IL6, which itself incorporates FedRAMP High, CNSSI 1253, and other frameworks; future revisions will explicitly map SL5 requirements to IL6.

Organized by NIST control family. "NIST Control Text" and "NIST Discussion" are taken verbatim from NIST SP 800-53 Rev 5. "SL5 Supplemental Guidance" and "Parameter Values" are additions specific to this standard. Where a Parameter Values section is absent or a specific assignment is not provided, organizations define values based on their specific context.

## 3.1 Access Control (AC)

### AC-3(2): Dual Authorization

**NIST Control Text:**

Enforce dual authorization for [Assignment: organization-defined privileged commands and/or other actions].

**NIST Discussion:**

Dual authorization, also known as two-person control, reduces risk related to insider threats. Dual authorization mechanisms require the approval of two authorized individuals to execute. To reduce the risk of collusion, organizations consider rotating dual authorization duties. Organizations consider the risk associated with implementing dual authorization mechanisms when immediate responses are necessary to ensure public and environmental safety.

**SL5 Supplemental Guidance:**

Organizations define privileged commands and actions requiring dual authorization based on their operational context, with emphasis on operations involving covered models or cryptographic keys protecting them. Dual authorization reduces single-actor compromise risk, which is particularly important given the limitations of private-sector vetting against sophisticated adversaries. At least one SenL-5 custodian must participate in dual authorization for weight operations [26].

Technical enforcement through cryptographic split-knowledge mechanisms or multi-party authorization workflows.

### AC-4: Information Flow Enforcement

**NIST Control Text:**

Enforce approved authorizations for controlling the flow of information within the system and between connected systems based on [Assignment: organization-defined information flow control policies].

**NIST Discussion:**

Information flow control regulates where information can travel within a system and between systems (in contrast to who is allowed to access the information) and without regard to subsequent accesses to that information. Flow control restrictions include blocking external traffic that claims to be from within the organization, keeping export-controlled information from being transmitted in the clear to the Internet, restricting web requests that are not from the internal web proxy server, and limiting information transfers between organizations based on data structures and content. Transferring information between organizations may require an agreement specifying how the information flow is enforced (see CA-3 ). Transferring information between systems in different security or privacy domains with different security or privacy policies introduces the risk that such transfers violate one or more domain security or privacy policies. In such situations, information owners/stewards provide guidance at designated policy enforcement points between connected systems. Organizations consider mandating specific architectural solutions to enforce specific security and privacy policies. Enforcement includes prohibiting information transfers between connected systems (i.e., allowing access only), verifying write permissions before accepting information from another security or privacy domain or connected system, employing hardware mechanisms to enforce one-way information flows, and implementing trustworthy regrading mechanisms to reassign security or privacy attributes and labels.

Organizations commonly employ information flow control policies and enforcement mechanisms to control the flow of information between designated sources and destinations within systems and between connected systems. Flow control is based on the characteristics of the information and/or the information path. Enforcement occurs, for example, in boundary protection devices that employ rule sets or establish configuration settings that restrict system services, provide a packet-filtering capability based on header information, or provide a message-filtering capability based on message content. Organizations also consider the trustworthiness of filtering and/or inspection mechanisms (i.e., hardware, firmware, and software components) that are critical to information flow enforcement. Control enhancements 3 through 32 primarily address cross-domain solution needs that focus on more advanced filtering techniques, in-depth analysis, and stronger flow enforcement mechanisms implemented in cross-domain products, such as high-assurance guards. Such capabilities are generally not available in commercial off-the-shelf products. Information flow enforcement also applies to control plane traffic (e.g., routing and DNS).

**Parameter Values:**

- Assignment (flow control policies): All external data must be isolated in staging area; transfer to internal systems permitted only for data that completes screening successfully; data that fails screening must be quarantined and prevented from transfer

**SL5 Supplemental Guidance:**

External data may contain adversarial content including backdoor attacks in training data, jailbreak attempts in inference inputs, and malicious patterns in code (including inputs designed to compromise LLM-based code review or hardening systems). As models handle critical security engineering tasks and sensitive AI research, successful attacks could lead to sabotaged research, exfiltration of covered models, or compromised security outputs. Correlated supply chain attacks targeting both software and AI training data could amplify impact.

The organization enforces mandatory staging isolation: all external data is received into staging areas physically separated from internal systems. Transfer mechanisms enforce screening requirements, preventing bypass of the isolation boundary. Data remains isolated until automated screening validates it. This architecture prevents unscreened or malicious data from reaching internal systems.

### AC-4(9): Human Reviews

**NIST Control Text:**

Enforce the use of human reviews for [Assignment: organization-defined information flows] under the following conditions: [Assignment: organization-defined conditions].

**NIST Discussion:**

Organizations define security or privacy policy filters for all situations where automated flow control decisions are possible. When a fully automated flow control decision is not possible, then a human review may be employed in lieu of or as a complement to automated security or privacy policy filtering. Human reviews may also be employed as deemed necessary by organizations.

**Parameter Values:**

- Assignment (information flows): Data quarantined by automated detection systems (AC-4(15))
- Assignment (conditions): Automated detection systems flag content as potentially malicious; detection systems encounter processing errors or ambiguous cases

**SL5 Supplemental Guidance:**

The organization requires human review of quarantined data before making clearance or rejection decisions. The organization defines review scope based on operational capacity and false positive rates: either all quarantined data or a sample sufficient to validate detection effectiveness.

Human reviews provide oversight of automated detection, enabling identification of false positives, confirmation of true detections, and discovery of attack patterns not yet captured by automated mechanisms. Higher-risk data receives more intensive review.

### AC-4(15): Detection of Unsanctioned Information

**NIST Control Text:**

When transferring information between different security domains, examine the information for the presence of [Assignment: organization-defined unsanctioned information] and prohibit the transfer of such information in accordance with the [Assignment: organization-defined organization-defined security or privacy policy].

**NIST Discussion:**

Unsanctioned information includes malicious code, information that is inappropriate for release from the source network, or executable code that could disrupt or harm the services or systems on the destination network.

**Parameter Values:**

- Assignment (unsanctioned information): Adversarial content (poisoning attempts, jailbreak attempts, adversarial examples)

**SL5 Supplemental Guidance:**

Adversarial content designed to compromise AI models or data center operations constitutes unsanctioned information. Detection applies to all data that could be consumed as input by AI models: training datasets, evaluation data, prompts, code, configuration files, images, audio, video, and structured data. All supported modalities (text, images, video, audio, code, structured data) must be screened regardless of intended use.

Organizations determine detection thresholds through risk assessment (per RA-3), balancing false positives against false negatives. Risk-based tiering applies stricter thresholds to higher-risk data (e.g., training data, infrastructure code).

Detection systems must resist adversarial bypass attacks. Robust adversarial content detection remains an open research problem; best-available defensive measures are insufficient against sophisticated adversaries. Organizations invest in research and development to advance detection capabilities, understanding that breakthroughs in adversarial robustness are necessary to fully address this threat. Section 1.3 highlights this as a key open question.

## 3.2 Configuration Management (CM)

### CM-7(5): Least Functionality | Authorized Software — Allow-by-exception

**NIST Control Text:**

a. Identify [Assignment: organization-defined software programs];

b. Employ a deny-all, permit-by-exception policy to allow the execution of authorized software programs on the system; and

c. Review and update the list of authorized software programs [Assignment: organization-defined frequency].

**NIST Discussion:**

Authorized software programs can be limited to specific versions or from a specific source. To facilitate a comprehensive authorized software process and increase the strength of protection for attacks that bypass application level authorized software, software programs may be decomposed into and monitored at different levels of detail. These levels include applications, application programming interfaces, application modules, scripts, system processes, system services, kernel functions, registries, drivers, and dynamic link libraries. The concept of permitting the execution of authorized software may also be applied to user actions, system ports and protocols, IP addresses/ranges, websites, and MAC addresses. Organizations consider verifying the integrity of authorized software programs using digital signatures, cryptographic checksums, or hash functions. Verification of authorized software can occur either prior to execution or at system startup. The identification of authorized URLs for websites is addressed in CA-3(5) and SC-7.

**SL5 Supplemental Guidance:**

Weight Enclaves employ deny-all, permit-by-exception policies ensuring only vetted code executes with direct access to covered models. Authorized software undergoes testing and evaluation appropriate to the nation-state threat model (see SA-11) and cryptographic signing per SC-12 before deployment. Interactive coding environments are prohibited within Weight Enclaves; all code must complete testing and signing before enclave deployment.

The broader SL5 Network (outside Weight Enclaves) supports rapid prototyping and experimentation by enabling the execution of code that has not passed the full human review process. A primary use case is automated AI R&D where covered models generate and execute research experiments; these experiments execute in confined physical or virtual machine environments with limited privileges in the broader SL5 Network, not within Weight Enclaves. Organizations apply allow-by-exception principles for software executing outside confined environments, balancing development velocity with security.

## 3.3 Identification and Authentication (IA)

### IA-3: Device Identification and Authentication

**NIST Control Text:**

Uniquely identify and authenticate [Assignment: organization-defined devices and/or types of devices] before establishing a [Selection (one or more): local; remote; network] connection.

**NIST Discussion:**

Devices that require unique device-to-device identification and authentication are defined by type, device, or a combination of type and device. Organization-defined device types include devices that are not owned by the organization. Systems use shared known information (e.g., Media Access Control [MAC], Transmission Control Protocol/Internet Protocol [TCP/IP] addresses) for device identification or organizational authentication solutions (e.g., Institute of Electrical and Electronics Engineers (IEEE) 802.1x and Extensible Authentication Protocol [EAP], RADIUS server with EAP-Transport Layer Security [TLS] authentication, Kerberos) to identify and authenticate devices on local and wide area networks. Organizations determine the required strength of authentication mechanisms based on the security categories of systems and mission or business requirements. Because of the challenges of implementing device authentication on a large scale, organizations can restrict the application of the control to a limited number/type of devices based on mission or business needs.

**Parameter Values:**

- Assignment (devices): AI accelerators within Weight Enclaves
- Selection: Local; remote; network connection

**SL5 Supplemental Guidance:**

AI accelerators authenticate using cryptographic mechanisms anchored in a hardware root of trust. For distributed operation, accelerators authenticate each other before exchanging data, without host-mediated trust.

Accelerators support remote attestation: cryptographically proving their identity and configuration state to remote parties. Attestation enables data providers to verify they are communicating with a legitimate accelerator running authorized firmware before sending sensitive data. Boot measurements collected during secure boot are signed using hardware-protected keys, creating a cryptographic proof that can be verified without trusting the host.

## 3.4 Physical and Environmental Protection (PE)

### PE-2(3): Restrict Unescorted Access

**NIST Control Text:**

Restrict unescorted access to the facility where the system resides to personnel with [Selection (one or more): security clearances for all information contained within the system; formal access authorizations for all information contained within the system; need for access to all information contained within the system; [Assignment: organization-defined physical access authorizations] ].

**NIST Discussion:**

Individuals without required security clearances, access approvals, or need to know are escorted by individuals with appropriate physical access authorizations to ensure that information is not exposed or otherwise compromised.

**Parameter Values:**

- Selection: Security clearances (SenL-5 Custodial clearance) [26]

**SL5 Supplemental Guidance:**

Only personnel with SenL-5 (Custodial) clearance may enter Red Zones unescorted, and two authorized individuals must be present at all times (two-person integrity). Anyone without SenL-5—including SenL-4 holders and vendors—requires continuous escort by a SenL-5 holder, with all activities logged [26].

---

### PE-3(8): Access Control Vestibules

**NIST Control Text:**

Employ access control vestibules at [Assignment: organization-defined locations].

**NIST Discussion:**

An access control vestibule is part of a physical access control system that typically provides a space between two sets of interlocking doors. Vestibules are designed to prevent unauthorized individuals from following authorized individuals into facilities with controlled access. This activity, also known as piggybacking or tailgating, results in unauthorized access to the facility. Interlocking door controllers can be used to limit the number of individuals who enter controlled access points and to provide containment areas while authorization for physical access is verified. Interlocking door controllers can be fully automated (i.e., controlling the opening and closing of the doors) or partially automated (i.e., using security guards to control the number of individuals entering the containment area).

**Parameter Values:**

- Assignment (locations): All Red Zone entry and exit points

**SL5 Supplemental Guidance:**

All Red Zone entry points use access control vestibules (mantraps) with interlocking doors, multi-factor authentication, and anti-tailgating sensors. Mantraps prevent tailgating—following an authorized person through a door—and credential sharing.

### PE-19(1): National Emissions Policies and Procedures

**NIST Control Text:**

Protect system components, associated data communications, and networks in accordance with national Emissions Security policies and procedures based on the security category or classification of the information.

**NIST Discussion:**

Emissions Security (EMSEC) policies include the former TEMPEST policies.

**SL5 Supplemental Guidance:**

Red Zones implement TEMPEST countermeasures per TEMPEST/1-92—RF shielding to prevent electromagnetic signal egress and dedicated power conditioning to prevent power-line analysis [9], [21]. See Section 2 ( ICD 705 Facility Requirements) for construction details.

## 3.5 Program Management (PM)

### PM-12: Insider Threat Program

**NIST Control Text:**

Implement an insider threat program that includes a cross-discipline insider threat incident handling team.

**NIST Discussion:**

Organizations that handle classified information are required, under Executive Order 13587 EO 13587 and the National Insider Threat Policy ODNI NITP , to establish insider threat programs. The same standards and guidelines that apply to insider threat programs in classified environments can also be employed effectively to improve the security of controlled unclassified and other information in non-national security systems. Insider threat programs include controls to detect and prevent malicious insider activity through the centralized integration and analysis of both technical and nontechnical information to identify potential insider threat concerns. A senior official is designated by the department or agency head as the responsible individual to implement and provide oversight for the program. In addition to the centralized integration and analysis capability, insider threat programs require organizations to prepare department or agency insider threat policies and implementation plans, conduct host-based user monitoring of individual employee activities on government-owned classified computers, provide insider threat awareness training to employees, receive access to information from offices in the department or agency for insider threat analysis, and conduct self-assessments of department or agency insider threat posture.

Insider threat programs can leverage the existence of incident handling teams that organizations may already have in place, such as computer security incident response teams. Human resources records are especially important in this effort, as there is compelling evidence to show that some types of insider crimes are often preceded by nontechnical behaviors in the workplace, including ongoing patterns of disgruntled behavior and conflicts with coworkers and other colleagues. These precursors can guide organizational officials in more focused, targeted monitoring efforts. However, the use of human resource records could raise significant concerns for privacy. The participation of a legal team, including consultation with the senior agency official for privacy, ensures that monitoring activities are performed in accordance with applicable laws, executive orders, directives, regulations, policies, standards, and guidelines.

**SL5 Supplemental Guidance:**

Continuous monitoring intensity scales with Sensitivity Level, from periodic criminal checks and account auditing (SenL-1/2) through intensive monitoring of compartmented system interactions (SenL-5). Monitoring techniques, trigger thresholds, and jurisdictional variations are specified in a separate SenL Framework Document [26].

The insider threat program coordinates information security, personnel security, legal, HR, and physical security. Escalation triggers protective actions including access suspension (per AC-2(13)) when risk indicators emerge.

## 3.6 Personnel Security (PS)

Personnel security requirements in this section reference the SenL Framework Document, a separate detailed specification covering tier definitions, vetting procedures, adjudication criteria, and operational safeguards.

## PS-2: Position Risk Designation

**NIST Control Text:**

a. Assign a risk designation to all organizational positions;

b. Establish screening criteria for individuals filling those positions; and

c. Review and update position risk designations [Assignment: organization-defined frequency].

**NIST Discussion:**

Position risk designations reflect Office of Personnel Management (OPM) policy and guidance. Proper position designation is the foundation of an effective and consistent suitability and personnel security program. The Position Designation System (PDS) assesses the duties and responsibilities of a position to determine the degree of potential damage to the efficiency or integrity of the service due to misconduct of an incumbent of a position and establishes the risk level of that position. The PDS assessment also determines if the duties and responsibilities of the position present the potential for position incumbents to bring about a material adverse effect on national security and the degree of that potential effect, which establishes the sensitivity level of a position. The results of the assessment determine what level of investigation is conducted for a position. Risk designations can guide and inform the types of authorizations that individuals receive when accessing organizational information and information systems. Position screening criteria include explicit information security role appointment requirements. Parts 1400 and 731 of Title 5, Code of Federal Regulations, establish the requirements for organizations to evaluate relevant covered positions for a position sensitivity and position risk designation commensurate with the duties and responsibilities of those positions.

**Parameter Values:**

- Assignment (frequency): Annually and upon significant changes to position duties or access requirements

**SL5 Supplemental Guidance:**

The organization assigns one of five Sensitivity Levels (SenL-1 through SenL-5) to every position based on access to covered models and security-critical infrastructure. Tiers range from baseline corporate access (SenL-1) through compartmented weight custodian roles (SenL-5). The SenL Framework Document specifies detailed tier definitions, access criteria, and edge case handling [26].

## PS-3: Personnel Screening

**NIST Control Text:**

a. Screen individuals prior to authorizing access to the system; and

b. Rescreen individuals in accordance with [Assignment: organization-defined organization-defined conditions requiring rescreening and, where rescreening is so indicated, the frequency of rescreening].

**NIST Discussion:**

Personnel screening and rescreening activities reflect applicable laws, executive orders, directives, regulations, policies, standards, guidelines, and specific criteria established for the risk designations of assigned positions. Examples of personnel screening include background investigations and agency checks. Organizations may define different rescreening conditions and frequencies for personnel accessing systems based on types of information processed, stored, or transmitted by the systems.

**Parameter Values:**

- Assignment (conditions/frequency):
    - Rescreening frequency increases with tier: every 18 months (SenL-1) to every 13 months (SenL-5) [26]
    - Rescreen upon role change requiring higher SenL or upon triggering events from continuous monitoring [26]

**SL5 Supplemental Guidance:**

All tiers use a “Private SF-86” disclosure packet—a comprehensive background disclosure form modeled on the government SF-86 but using lawfully accessible private-sector sources. Higher tiers (SenL-4/5) add verified subject and reference interviews for character and reliability assessment. Private-sector vetting is inherently limited; organizations may need to pursue government involvement (through classified contract pathways, information sharing arrangements, or new statutory authority) to achieve adequate personnel security for the stated threat model [26].

Provisional access during vetting requires compensating controls as specified in the SenL Framework Document [26].

### PS-6: Access Agreements

**NIST Control Text:**

a. Develop and document access agreements for organizational systems;

b. Review and update the access agreements [Assignment: organization-defined frequency] ; and

c. Verify that individuals requiring access to organizational information and systems:

1. Sign appropriate access agreements prior to being granted access; and

2. Re-sign access agreements to maintain access to organizational systems when access agreements have been updated or [Assignment: organization-defined frequency].

**NIST Discussion:**

Access agreements include nondisclosure agreements, acceptable use agreements, rules of behavior, and conflict-of-interest agreements. Signed access agreements include an acknowledgement that individuals have read, understand, and agree to abide by the constraints associated with organizational systems to which access is authorized. Organizations can use electronic signatures to acknowledge access agreements unless specifically prohibited by organizational policy.

**Parameter Values:**

- Assignment (frequency - review/update): Annually
- Assignment (frequency - re-sign): Upon SenL designation change, upon rescreening per PS-3, when agreements are updated, or annually

**SL5 Supplemental Guidance:**

Access agreements scale with Sensitivity Level. SenL-1/2 include standard NDA and monitoring acknowledgment. Higher tiers add progressively stricter obligations: foreign contact reporting and secondary employment restrictions (SenL-3), dual authorization acknowledgment and travel notification (SenL-4), custodian-specific protocols and post-employment restrictions (SenL-5). Complete tier-specific requirements are in the SenL Framework Document [26].

## 3.7 System and Services Acquisition (SA)

### SA-4: Acquisition Process

**NIST Control Text:**

Include the following requirements, descriptions, and criteria, explicitly or by reference, using [Selection (one or more): standardized contract language; [Assignment: organization-defined contract language] ] in the acquisition contract for the system, system component, or system service:

a. Security and privacy functional requirements;

b. Strength of mechanism requirements;

c. Security and privacy assurance requirements;

d. Controls needed to satisfy the security and privacy requirements.

e. Security and privacy documentation requirements;

f. Requirements for protecting security and privacy documentation;

g. Description of the system development environment and environment in which the system is intended to operate;

h. Allocation of responsibility or identification of parties responsible for information security, privacy, and supply chain risk management; and

i. Acceptance criteria.

**NIST Discussion:**

Security and privacy functional requirements are typically derived from the high-level security and privacy requirements described in SA-2 . The derived requirements include security and privacy capabilities, functions, and mechanisms. Strength requirements associated with such capabilities, functions, and mechanisms include degree of correctness, completeness, resistance to tampering or bypass, and resistance to direct attack. Assurance requirements include development processes, procedures, and methodologies as well as the evidence from development and assessment activities that provide grounds for confidence that the required functionality is implemented and possesses the required strength of mechanism. SP 800-160-1 describes the process of requirements engineering as part of the system development life cycle.

Controls can be viewed as descriptions of the safeguards and protection capabilities appropriate for achieving the particular security and privacy objectives of the organization and for reflecting the security and privacy requirements of stakeholders. Controls are selected and implemented in order to satisfy system requirements and include developer and organizational responsibilities.

Controls can include technical, administrative, and physical aspects. In some cases, the selection and implementation of a control may necessitate additional specification by the organization in the form of derived requirements or instantiated control parameter values. The derived requirements and control parameter values may be necessary to provide the appropriate level of implementation detail for controls within the system development life cycle.

Security and privacy documentation requirements address all stages of the system development life cycle. Documentation provides user and administrator guidance for the implementation and operation of controls. The level of detail required in such documentation is based on the security categorization or classification level of the system and the degree to which organizations depend on the capabilities, functions, or mechanisms to meet risk response expectations. Requirements can include mandated configuration settings that specify allowed functions, ports, protocols, and services. Acceptance criteria for systems, system components, and system services are defined in the same manner as the criteria for any organizational acquisition or procurement.

Organizations can determine other requirements that support security and operations, to include responsibilities for the organization and developer, and notification and timing requirements for support, maintenance and updates.

**SL5 Supplemental Guidance:**

**AI Accelerator Security Features:** Acquisition contracts for AI accelerators specify hardware security features as functional requirements: integrated root-of-trust with hardware-provisioned identity (SC-28(3)), device-controlled memory isolation (SC-49), interconnect encryption (SC-8(1)), execution integrity verification (SI-7(9), SI-7(10), SI-7(15)), attestation capability (IA-3), and tamper protection (SR-9, SR-9(1)).

These features require advance engagement with chip providers—security features must be designed into silicon before tapeout. Organizations communicate requirements during architecture and design phases, not after chips are available, and include provisions for pre-acceptance assessment (SR-5(2)).

**Shielded Rack Enclosures:** Nation-state adversaries can extract sensitive information from electromagnetic emissions. Advanced AI models could potentially modulate EM emissions for covert communication. Facility-level shielding (per Section 2) protects the perimeter; shielded rack enclosures provide defense in depth or internal compartmentalization.

Shielded rack enclosures meeting NSA 94-106 attenuation specifications (100dB from 1kHz to 10GHz) [10] or equivalent are required for systems within Weight Enclaves processing covered models. Organizations may deploy shielded racks for other systems based on threat modeling.

### SA-11: Developer Testing and Evaluation

**NIST Control Text:**

Require the developer of the system, system component, or system service, at all post-design stages of the system development life cycle, to:

a. Develop and implement a plan for ongoing security and privacy control assessments;

b. Perform [Selection (one or more): unit; integration; system; regression] testing/evaluation [Assignment: organization-defined frequency to conduct] at [Assignment: organization-defined depth and coverage];

c. Produce evidence of the execution of the assessment plan and the results of the testing and evaluation;

d. Implement a verifiable flaw remediation process; and

e. Correct flaws identified during testing and evaluation.

**NIST Discussion:**

Developmental testing and evaluation confirms that the required controls are implemented correctly, operating as intended, enforcing the desired security and privacy policies, and meeting established security and privacy requirements. Security properties of systems and the privacy of individuals may be affected by the interconnection of system components or changes to those components. The interconnections or changes—including upgrading or replacing applications, operating systems, and firmware—may adversely affect previously implemented controls. Ongoing assessment during development allows for additional types of testing and evaluation that developers can conduct to reduce or eliminate potential flaws. Testing custom software applications may require approaches such as manual code review, security architecture review, and penetration testing, as well as and static analysis, dynamic analysis, binary analysis, or a hybrid of the three analysis approaches.

Developers can use the analysis approaches, along with security instrumentation and fuzzing, in a variety of tools and in source code reviews. The security and privacy assessment plans include the specific activities that developers plan to carry out, including the types of analyses, testing, evaluation, and reviews of software and firmware components; the degree of rigor to be applied; the frequency of the ongoing testing and evaluation; and the types of artifacts produced during those processes. The depth of testing and evaluation refers to the rigor and level of detail associated with the assessment process. The coverage of testing and evaluation refers to the scope (i.e., number and type) of the artifacts included in the assessment process. Contracts specify the acceptance criteria for security and privacy assessment plans, flaw remediation processes, and the evidence that the plans and processes have been diligently applied. Methods for reviewing and protecting assessment plans, evidence, and documentation are commensurate with the security category or classification level of the system. Contracts may specify protection requirements for documentation.

**SL5 Supplemental Guidance:**

Apply SP 800-161 Rev 1 guidance [3]. Testing includes counterfeit detection, verification of component origins, examination of configuration settings, and physical inspection.

## SA-17: Developer Security and Privacy Architecture and Design

**NIST Control Text:**

Require the developer of the system, system component, or system service to produce a design specification and security and privacy architecture that:

a. Is consistent with the organization’s security and privacy architecture that is an integral part the organization’s enterprise architecture;

b. Accurately and completely describes the required security and privacy functionality, and the allocation of controls among physical and logical components; and

c. Expresses how individual security and privacy functions, mechanisms, and services work together to provide required security and privacy capabilities and a unified approach to protection.

**NIST Discussion:**

Developer security and privacy architecture and design are directed at external developers, although they could also be applied to internal (in-house) development. In contrast, PL-8 is directed at internal developers to ensure that organizations develop a security and privacy architecture that is integrated with the enterprise architecture. The distinction between SA-17 and PL-8 is especially important when organizations outsource the development of systems, system components, or system services and when there is a requirement to demonstrate consistency with the enterprise architecture and security and privacy architecture of the organization. ISO 15408-2, ISO 15408-3 , and SP 800-160-1 provide information on security architecture and design, including formal policy models, security-relevant components, formal and informal correspondence, conceptually simple design, and structuring for least privilege and testing.

**SL5 Supplemental Guidance:**

Apply SP 800-161 Rev 1 guidance [3]. Architecture decisions should address security and resilience considerations, including component selection strategies that enable availability through multiple supplier options.

### SA-20: Customized Development of Critical Components

**NIST Control Text:**

Reimplement or custom develop the following critical system components: [Assignment: organization-defined critical system].

**NIST Discussion:**

Organizations determine that certain system components likely cannot be trusted due to specific threats to and vulnerabilities in those components for which there are no viable security controls to adequately mitigate risk. Reimplementation or custom development of such components may satisfy requirements for higher assurance and is carried out by initiating changes to system components (including hardware, software, and firmware) such that the standard attacks by adversaries are less likely to succeed. In situations where no alternative sourcing is available and organizations choose not to reimplement or custom develop critical system components, additional controls can be employed. Controls include enhanced auditing, restrictions on source code and system utility access, and protection from deletion of system and application files.

**SL5 Supplemental Guidance:**

Apply SP 800-161 Rev 1 guidance [3] for customized development of critical components where supply chain risk is unacceptable, including maintaining source code, build scripts, and tests to ensure continued maintenance capability.

### SA-21: Developer Screening

**NIST Control Text:**

Require that the developer of [Assignment: organization-defined system, systems component, or system service]:

a. Has appropriate access authorizations as determined by assigned [Assignment: organization-defined official government duties] ; and

b. Satisfies the following additional personnel screening criteria: [Assignment: organization-defined additional personnel screening criteria].

**NIST Discussion:**

Developer screening is directed at external developers. Internal developer screening is addressed by PS-3 . Because the system, system component, or system service may be used in critical activities essential to the national or economic security interests of the United States, organizations have a strong interest in ensuring that developers are trustworthy. The degree of trust required of developers may need to be consistent with that of the individuals who access the systems, system components, or system services once deployed. Authorization and personnel screening criteria include clearances, background checks, citizenship, and nationality. Developer trustworthiness may also include a review and analysis of company ownership and relationships that the company has with entities that may potentially affect the quality and reliability of the systems, components, or services being developed. Satisfying the required access authorizations and personnel screening criteria includes providing a list of all individuals who are authorized to perform development activities on the selected system, system component, or system service so that organizations can validate that the developer has satisfied the authorization and screening requirements.

**SL5 Supplemental Guidance:**

Apply SP 800-161 Rev 1 guidance [3] for screening personnel with access to hardware development environments or critical components, including validation processes for both internal developers and key personnel from system integrators. External developer personnel are assigned Sensitivity Levels per the SenL Framework Document [26].

## 3.8 Supply Chain Risk Management (SR)

### SR-3(1): Diverse Supply Base

**NIST Control Text:**

Employ a diverse set of sources for the following system components and services: [Assignment: organization-defined organization-defined system components and services].

**NIST Discussion:**

Diversifying the supply of systems, system components, and services can reduce the probability that adversaries will successfully identify and target the supply chain and can reduce the impact of a supply chain event or compromise. Identifying multiple suppliers for replacement components can reduce the probability that the replacement component will become unavailable. Employing a diverse set of developers or logistics service providers can reduce the impact of a natural disaster or other supply chain event. Organizations consider designing the system to include diverse materials and components.

**SL5 Supplemental Guidance:**

Apply SP 800-161 Rev 1 guidance [3] for diversifying the supply base to eliminate single points of failure, particularly for critical components where feasible. Where market constraints limit supplier diversity, apply compensating supply chain risk mitigations.

### SR-3(3): Sub-Tier Flow Down

**NIST Control Text:**

Ensure that the controls included in the contracts of prime contractors are also included in the contracts of subcontractors.

**NIST Discussion:**

To manage supply chain risk effectively and holistically, it is important that organizations ensure that supply chain risk management controls are included at all tiers in the supply chain. This includes ensuring that Tier 1 (prime) contractors have implemented processes to facilitate the "flow down" of supply chain risk management controls to sub-tier contractors. The controls subject to flow down are identified in SR-3b.

**SL5 Supplemental Guidance:**

Apply SP 800-161 Rev 1 guidance [3] for contractual flow-down of security requirements from prime contractors to relevant sub-tier contractors throughout the supply chain, with due diligence on upstream dependencies including fourth- and fifth-party suppliers.

### SR-5(2): Assessments Prior to Selection, Acceptance, Modification, or Update

**NIST Control Text:**

Assess the system, system component, or system service prior to selection, acceptance, modification, or update.

**NIST Discussion:**

Organizational personnel or independent, external entities conduct assessments of systems, components, products, tools, and services to uncover evidence of tampering, unintentional and intentional vulnerabilities, or evidence of non-compliance with supply chain controls. These include malicious code, malicious processes, defective software, backdoors, and counterfeits. Assessments can include evaluations; design proposal reviews; visual or physical inspection; static and dynamic analyses; visual, x-ray, or magnetic particle inspections; simulations; white, gray, or black box testing; fuzz testing; stress testing; and penetration testing (see SR-6(1) ). Evidence generated during assessments is documented for follow-on actions by organizations. The evidence generated during the organizational or independent assessments of supply chain elements may be used to improve supply chain processes and inform the supply chain risk management process. The evidence can be leveraged in follow-on assessments. Evidence and other documentation may be shared in accordance with organizational agreements.

**SL5 Supplemental Guidance:**

Organizations assess a sample of AI accelerators before acceptance to verify that hardware security features function as specified: that memory isolation prevents host access, attestation produces valid cryptographic proofs, and firmware verification rejects unsigned code.

### SR-6: Supplier Assessments and Reviews

**NIST Control Text:**

Assess and review the supply chain-related risks associated with suppliers or contractors and the system, system component, or system service they provide [Assignment: organization-defined frequency].

**NIST Discussion:**

An assessment and review of supplier risk includes security and supply chain risk management processes, foreign ownership, control or influence (FOCI), and the ability of the supplier to effectively assess subordinate second-tier and third-tier suppliers and contractors. The reviews may be conducted by the organization or by an independent third party. The reviews consider documented processes, documented controls, all-source intelligence, and publicly available information related to the supplier or contractor. Organizations can use open-source information to monitor for indications of stolen information, poor development and quality control practices, information spillage, or counterfeits. In some cases, it may be appropriate or required to share assessment and review results with other organizations in accordance with any applicable rules, policies, or inter-organizational agreements or contracts.

**SL5 Supplemental Guidance:**

Apply SP 800-161 Rev 1 guidance [3] for rigorous, continuous assessment of all suppliers against consistent baseline criteria evaluating security, integrity, resilience, quality, trustworthiness, and authenticity.

### SR-9: Tamper Resistance and Detection

**NIST Control Text:**

Implement a tamper protection program for the system, system component, or system service.

**NIST Discussion:**

Anti-tamper technologies, tools, and techniques provide a level of protection for systems, system components, and services against many threats, including reverse engineering, modification, and substitution. Strong identification combined with tamper resistance and/or tamper detection is essential to protecting systems and components during distribution and when in use.

**SL5 Supplemental Guidance:**

AI accelerators within Weight Enclaves implement comprehensive tamper protection. Sensitive data must exist unencrypted during computation, making the compute cores attractive targets for physical attack. Tamper protection extends to all contexts where confidential data exists in plaintext—not just the root-of-trust components.

Specific mechanisms are determined based on threat model and risk assessment. Detection mechanisms identify tampering attempts; response mechanisms may include zeroization of sensitive data.

### SR-9(1): Multiple Stages of System Development Life Cycle

**NIST Control Text:**

Employ anti-tamper technologies, tools, and techniques throughout the system development life cycle.

**NIST Discussion:**

The system development life cycle includes research and development, design, manufacturing, acquisition, delivery, integration, operations and maintenance, and disposal. Organizations use a combination of hardware and software techniques for tamper resistance and detection. Organizations use obfuscation and self-checking to make reverse engineering and modifications more difficult, time-consuming, and expensive for adversaries. The customization of systems and system components can make substitutions easier to detect and therefore limit damage.

**SL5 Supplemental Guidance:**

Chip providers employ anti-tamper technologies throughout the accelerator development lifecycle, including design, manufacturing, and integration phases. An attacker who can modify the chip during manufacturing can install hardware backdoors that bypass all other security mechanisms.

### SR-10: Inspection of Systems or Components

**NIST Control Text:**

Inspect the following systems or system components [Selection (one or more): at random; at [Assignment: organization-defined frequency] ; upon [Assignment: organization-defined indications of need for inspection] ] to detect tampering: [Assignment: organization-defined systems or system components].

**NIST Discussion:**

The inspection of systems or systems components for tamper resistance and detection addresses physical and logical tampering and is applied to systems and system components removed from organization-controlled areas. Indications of a need for inspection include changes in packaging, specifications, factory location, or entity in which the part is purchased, and when individuals return from travel to high-risk locations.

**SL5 Supplemental Guidance:**

Apply SP 800-161 Rev 1 guidance [3] for physical inspection of critical hardware components prior to initial use and periodically thereafter, using techniques such as radiographic examination, material analysis, and electrical testing.

### SR-11: Component Authenticity

**NIST Control Text:**

a. Develop and implement anti-counterfeit policy and procedures that include the means to detect and prevent counterfeit components from entering the system; and

b. Report counterfeit system components to [Selection (one or more): source of counterfeit component; [Assignment: organization-defined external reporting organizations] ; [Assignment: organization-defined personnel or roles] ].

**NIST Discussion:**

Sources of counterfeit components include manufacturers, developers, vendors, and contractors. Anti-counterfeiting policies and procedures support tamper resistance and provide a level of protection against the introduction of malicious code. External reporting organizations include CISA.

**SL5 Supplemental Guidance:**

Apply SP 800-161 Rev 1 guidance [3] for prevention of counterfeit components through use of qualified bidders lists (QBL) and qualified manufacturers lists (QML). While not available to private companies unless contracting for the government, the Trusted Foundry program could be useful for some components if government cooperation is available [27].

### SR-13: Supplier Inventory *(SP 800-161)*

**NIST Control Text:**

a. Develop, document, and maintain an inventory of suppliers that:

1. Accurately and minimally reflects the organization's tier one suppliers that may present a cybersecurity risk in the supply chain;
2. Is at the level of granularity deemed necessary for assessing criticality and supply chain risk, tracking, and reporting;
3. Documents the following information for each tier one supplier: (i) unique identifier for procurement instrument; (ii) description of the supplied products and/or services; (iii) program, project, and/or system that uses the supplier's products and/or services; and (iv) assigned criticality level that aligns to the criticality of the program, project, and/or system.

b. Review and update the supplier inventory [Assignment: enterprise-defined frequency].

**SL5 Supplemental Guidance:**

Apply SP 800-161 Rev 1 guidance [3] for maintaining a comprehensive, criticality-based inventory of all suppliers documenting supplier identities, products provided, and assigned risk levels.

## 3.9 System and Communications Protection (SC)

### SC-7: Boundary Protection

**NIST Control Text:**

a. Monitor and control communications at the external managed interfaces to the system and at key internal managed interfaces within the system;

b. Implement subnetworks for publicly accessible system components that are [Selection: physically; logically] separated from internal organizational networks; and

c. Connect to external networks or systems only through managed interfaces consisting of boundary protection devices arranged in accordance with an organizational security and privacy architecture.

**NIST Discussion:**

Managed interfaces include gateways, routers, firewalls, guards, network-based malicious code analysis, virtualization systems, or encrypted tunnels implemented within a security architecture. Subnetworks that are physically or logically separated from internal networks are referred to as demilitarized zones or DMZs. Restricting or prohibiting interfaces within organizational systems includes restricting external web traffic to designated web servers within managed interfaces, prohibiting external traffic that appears to be spoofing internal addresses, and prohibiting internal traffic that appears to be spoofing external addresses. SP 800-189 provides additional information on source address validation techniques to prevent ingress and egress of traffic with spoofed addresses. Commercial telecommunications services are provided by network components and consolidated management systems shared by customers. These services may also include third party-provided access lines and other service elements. Such services may represent sources of increased risk despite contract security provisions. Boundary protection may be implemented as a common control for all or part of an organizational network such that the boundary to be protected is greater than a system-specific boundary (i.e., an authorization boundary).

**Parameter Values:**

- Selection (subnetworks): Not applicable - SL5 Network contains no publicly accessible system components

**SL5 Supplemental Guidance:**

The SL5 Network encompasses SL5 model development, training, and deployment operations. External network connections are prohibited to prevent unauthorized access and exfiltration while supporting SL5 activities.

The SL5 Network may span multiple physical locations connected through managed interfaces with encrypted channels. Network access locations implement physical security per PE family controls.

### SC-7(10): Prevent Exfiltration

**NIST Control Text:**

a. Prevent the exfiltration of information; and

b. Conduct exfiltration tests [Assignment: organization-defined frequency].

**NIST Discussion:**

Prevention of exfiltration applies to both the intentional and unintentional exfiltration of information. Techniques used to prevent the exfiltration of information from systems may be implemented at internal endpoints, external boundaries, and across managed interfaces and include adherence to protocol formats, monitoring for beaconing activity from systems, disconnecting external network interfaces except when explicitly needed, employing traffic profile analysis to detect deviations from the volume and types of traffic expected, call backs to command and control centers, conducting penetration testing, monitoring for steganography, disassembling and reassembling packet headers, and using data loss and data leakage prevention tools. Devices that enforce strict adherence to protocol formats include deep packet inspection firewalls and Extensible Markup Language (XML) gateways. The devices verify adherence to protocol formats and specifications at the application layer and identify vulnerabilities that cannot be detected by devices that operate at the network or transport layers. The prevention of exfiltration is similar to data loss prevention or data leakage prevention and is closely associated with cross-domain solutions and system guards that enforce information flow requirements.

**SL5 Supplemental Guidance:**

Organizations prevent exfiltration of covered models through physical bandwidth limitation on outbound flows from Weight Enclaves. Hardware-enforced rate limiting provides deterministic throughput caps that prevent weight exfiltration, even if attempts go undetected. This assumes covered models are substantially larger than required outputs crossing the boundary; if this assumption does not hold, alternative controls are required. Limits are calibrated based on model size and organizational threat model to make weight exfiltration infeasible within acceptable time scales.

Boundary bandwidth may be preserved through multiple strategies: placing systems requiring high data volumes within the Weight Enclave per SC-7(21) to avoid boundary bandwidth consumption entirely, or minimizing outbound data volume by processing outputs within the Weight Enclave before transfer. Examples include stripping thinking tokens from inference responses, aggregating results, extracting only necessary information, or exfiltrating high-level experiment plans to be implemented by less capable models running in the broader SL5 Network.

Organizations implement additional exfiltration prevention mechanisms, such as bandwidth accounting and monitoring, detection and response capabilities per SI-4 and IR-4, monitoring for steganography, and traffic profile analysis. Physical bandwidth limitation serves as the hard cap that bounds exfiltration risk even if other mechanisms are bypassed or detection fails.

## SC-7(21): Isolation of System Components

**NIST Control Text:**

Employ boundary protection mechanisms to isolate [Assignment: organization-defined system components] supporting [Assignment: organization-defined missions and/or business functions].

**NIST Discussion:**

Organizations can isolate system components that perform different mission or business functions. Such isolation limits unauthorized information flows among system components and provides the opportunity to deploy greater levels of protection for selected system components. Isolating system components with boundary protection mechanisms provides the capability for increased protection of individual system components and to more effectively control information flows between those components. Isolating system components provides enhanced protection that limits the potential harm from hostile cyber-attacks and errors. The degree of isolation varies depending upon the mechanisms chosen. Boundary protection mechanisms include routers, gateways, and firewalls that separate system components into physically separate networks or subnetworks; cross-domain devices that separate subnetworks; virtualization techniques; and the encryption of information flows among system components using distinct encryption keys.

**Parameter Values:**

- Assignment (system components): Weight Enclaves (air-gapped subnetworks hosting covered models and weight-accessing systems)
- Assignment (missions/business functions): Model training, inference, fine-tuning, mechanistic interpretability, and other operations requiring direct weight access

**SL5 Supplemental Guidance:**

Weight Enclaves isolate systems requiring direct access to covered models within the SL5 Network. This isolation protects against weight exfiltration while enabling operations such as training, inference, fine-tuning, and mechanistic interpretability. Boundary protection mechanisms enable code deployment and API access while preventing weight exfiltration.

Each Weight Enclave resides in a single physical facility. Multiple Weight Enclaves may be established at different facilities, but may not be directly connected via network. Transfers exceeding the Weight Enclave outbound bandwidth limit must occur via encrypted physical media with appropriate physical safeguards. This has significant implications for geographically distributed training; Section 1.3 highlights this as a key open question.

Systems not requiring direct weight access operate outside Weight Enclaves, including lower-risk models. Weight Enclaves execute only authorized software (CM-7(5)).

## SC-8(1): Cryptographic Protection

**NIST Control Text:**

Implement cryptographic mechanisms to [Selection (one or more): prevent unauthorized disclosure of information; detect changes to information] during transmission.

**NIST Discussion:**

Encryption protects information from unauthorized disclosure and modification during transmission. Cryptographic mechanisms that protect the confidentiality and integrity of information during transmission include TLS and IPSec. Cryptographic mechanisms used to protect information integrity include cryptographic hash functions that have applications in digital signatures, checksums, and message authentication codes.

**Parameter Values:**

- Selection: Prevent unauthorized disclosure; detect changes

**SL5 Supplemental Guidance:**

**Accelerator Interconnect Encryption:** AI accelerators within Weight Enclaves cryptographically protect all data transmitted over chip-to-chip interconnects (e.g., NVLink, UALink, custom fabrics). Hardware-level encryption prevents interception from physical interconnects during distributed operation.

Interconnect encryption protects data between accelerators; end-to-end encryption (where data arrives encrypted from origin and is re-encrypted before host-accessible export) protects data from host access at trust boundary crossings. Both are required. The accelerator must revoke host memory access before decrypting incoming data and complete encryption before restoring host access.

**Inter-Facility Encryption:** The organization implements cryptographic protection for all inter-facility network traffic within the SL5 Network using inline network encryptors deployed at facility boundaries. Inline network encryptors must use cryptographic modules validated to FIPS 140-3 Level 3 minimum [11]. This is aligned with FedRAMP's requirement to use FIPS-validated cryptography and is intentionally stronger on validation level [5], [21]. The organization deploys at least two inline network encryptors from different suppliers in series at each inter-facility connection per SC-29.

### SC-8(5): Protected Distribution System

**NIST Control Text:**

Implement [Assignment: organization-defined protected distribution system] to [Selection (one or more): prevent unauthorized disclosure of information; detect changes to information] during transmission.

**NIST Discussion:**

The purpose of a protected distribution system is to deter, detect, and/or make difficult physical access to the communication lines that carry national security information.

**Parameter Values:**

- Assignment (PDS): Protected Distribution System per CNSSI 7003 [8]
- Selection: Prevent unauthorized disclosure; detect changes to information

**SL5 Supplemental Guidance:**

All Weight Enclave network traffic leaving the Red Zone perimeter requires PDS per CNSSI 7003. Unlike standard SCIF requirements (which apply only to unencrypted traffic), SL5 requires PDS for all Weight Enclave traffic including encrypted inter-building connections.

### SC-13: Cryptographic Protection

**NIST Control Text:**

a. Determine the [Assignment: organization-defined cryptographic uses] ; and

b. Implement the following types of cryptography required for each specified cryptographic use: [Assignment: organization-defined types of cryptography].

**NIST Discussion:**

Cryptography can be employed to support a variety of security solutions, including the protection of classified information and controlled unclassified information, the provision and implementation of digital signatures, and the enforcement of information separation when authorized individuals have the necessary clearances but lack the necessary formal access approvals. Cryptography can also be used to support random number and hash generation. Generally applicable cryptographic standards include FIPS-validated cryptography and NSA-approved cryptography. For example, organizations that need to protect classified information may specify the use of NSA-approved cryptography. Organizations that need to provision and implement digital signatures may specify the use of FIPS-validated cryptography. Cryptography is implemented in accordance with applicable laws, executive orders, directives, regulations, policies, standards, and guidelines.

**SL5 Supplemental Guidance:**

This standard specifies FIPS 140-3 Level 3 module validation for inline network encryptors at inter-facility boundaries per SC-8(1). Cryptographic uses and types for other contexts are specified by applicable frameworks (CNSSI 1253 specifies FIPS-validated cryptography for all components that support it [4]) and organizational policies. Government and government contractors have access to NSA Type 1 Certified network encryptors, which is a higher standard than FIPS 140-3 Level 3 [28]. If government cooperation is available, these may be preferable.

### SC-15(3): Disabling and Removal in Secure Work Areas

**NIST Control Text:**

Disable or remove collaborative computing devices and applications from [Assignment: organization-defined systems or system components] in [Assignment: organization-defined secure work areas].

**NIST Discussion:**

Failing to disable or remove collaborative computing devices and applications from systems or system components can result in compromises of information, including eavesdropping on conversations. A Sensitive Compartmented Information Facility (SCIF) is an example of a secure work area.

**Parameter Values:**

- Assignment (systems): All systems and devices
- Assignment (secure work areas): Red Zones

**SL5 Supplemental Guidance:**

No wireless devices or collaborative computing devices (cameras, microphones, video conferencing) are permitted in Red Zones. All equipment must be hardwired with wireless capabilities physically removed or disabled.

### SC-28(3): Cryptographic Keys

**NIST Control Text:**

Provide protected storage for cryptographic keys [Selection: [Assignment: organization-defined safeguards] ; hardware-protected key store].

**NIST Discussion:**

A Trusted Platform Module (TPM) is an example of a hardware-protected data store that can be used to protect cryptographic keys.

**Parameter Values:**

- Selection: Hardware-protected key store

**SL5 Supplemental Guidance:**

AI accelerators within Weight Enclaves provide a dedicated secure element for cryptographic keys used in encrypted data paths and attestation. The host system cannot access this key storage.

Hardware-provisioned identity keys (e.g., e-fuse keys embedded during manufacturing) serve as the accelerator's unforgeable identity, enabling the accelerator to prove its identity to remote parties.

### SC-29: Heterogeneity

**NIST Control Text:**

Employ a diverse set of information technologies for the following system components in the implementation of the system: [Assignment: organization-defined system components].

**NIST Discussion:**

Increasing the diversity of information technologies within organizational systems reduces the impact of potential exploitations or compromises of specific technologies. Such diversity protects against common mode failures, including those failures induced by supply chain attacks. Diversity in information technologies also reduces the likelihood that the means adversaries use to compromise one system component will be effective against other system components, thus further increasing the adversary work factor to successfully complete planned attacks. An increase in diversity may add complexity and management overhead that could ultimately lead to mistakes and unauthorized configurations.

**Parameter Values:**

- Assignment (system components): Inline network encryptors at facility boundaries (at least two encryptors from different suppliers per inter-facility connection)

**SL5 Supplemental Guidance:**

The organization deploys at least two inline network encryptors from different suppliers in series for each inter-facility connection, consistent with the NSA "Rule of Two" [14]. Different suppliers means different manufacturers or companies producing the encryptors. This heterogeneity protects against supplier-specific implementation vulnerabilities in firmware, hardware design, key management implementations, or protocol handling.

### SC-32: System Partitioning

**NIST Control Text:**

Partition the system into [Assignment: organization-defined system components] residing in separate [Selection: physical; logical] domains or environments based on [Assignment: organization-defined circumstances for the physical or logical separation of components].

**NIST Discussion:**

System partitioning is part of a defense-in-depth protection strategy. Organizations determine the degree of physical separation of system components. Physical separation options include physically distinct components in separate racks in the same room, critical components in separate rooms, and geographical separation of critical components. Security categorization can guide the selection of candidates for domain partitioning. Managed interfaces restrict or prohibit network access and information flow among partitioned system components.

**SL5 Supplemental Guidance:**

Weight Enclaves constitute separate physical and logical domains within the SL5 Network. This partitioning protects covered models from unauthorized access by other SL5 Network components while enabling interactions through managed interfaces per SC-7(21). Separation is enforced through network segmentation, access controls, and physical isolation.

### SC-49: Hardware-Enforced Separation and Policy Enforcement

**NIST Control Text:**

Implement hardware-enforced separation and policy enforcement mechanisms between [Assignment: organization-defined security domains].

**NIST Discussion:**

System owners may require additional strength of mechanism and robustness to ensure domain separation and policy enforcement for specific types of threats and environments of operation. Hardware-enforced separation and policy enforcement provide greater strength of mechanism than software-enforced separation and policy enforcement.

**Parameter Values:**

- Assignment (security domains): AI accelerator secure execution environment; host system

**SL5 Supplemental Guidance:**

AI accelerators within Weight Enclaves implement hardware-enforced separation establishing the accelerator as an independent security domain from the host. The accelerator prevents memory access from the host to regions containing sensitive data via DMA or other mechanism. This configuration is controlled by the accelerator and not the host. The host cannot override these policies even with privileged access.

## 3.10 System and Information Integrity (SI)

### SI-3: Malicious Code Protection

**NIST Control Text:**

a. Implement [Selection (one or more): signature-based; non-signature-based] malicious code protection mechanisms at system entry and exit points to detect and eradicate malicious code;

b. Automatically update malicious code protection mechanisms as new releases are available in accordance with organizational configuration management policy and procedures;

c. Configure malicious code protection mechanisms to:

1. Perform periodic scans of the system [Assignment: organization-defined frequency] and real-time scans of files from external sources at [Selection (one or more): endpoint; network entry and exit points] as the files are downloaded, opened, or executed in accordance with organizational policy; and

2. [Selection (one or more): block malicious code; quarantine malicious code; take [Assignment: organization-defined action] ] ; and send alert to [Assignment: organization-defined personnel or roles] in response to malicious code detection; and

d. Address the receipt of false positives during malicious code detection and eradication and the resulting potential impact on the availability of the system.

**NIST Discussion:**

System entry and exit points include firewalls, remote access servers, workstations, electronic mail servers, web servers, proxy servers, notebook computers, and mobile devices. Malicious code includes viruses, worms, Trojan horses, and spyware. Malicious code can also be encoded in various formats contained within compressed or hidden files or hidden in files using techniques such as steganography. Malicious code can be inserted into systems in a variety of ways, including by electronic mail, the world-wide web, and portable storage devices. Malicious code insertions occur through the exploitation of system vulnerabilities. A variety of technologies and methods exist to limit or eliminate the effects of malicious code.

Malicious code protection mechanisms include both signature- and nonsignature-based technologies. Nonsignature-based detection mechanisms include artificial intelligence techniques that use heuristics to detect, analyze, and describe the characteristics or behavior of malicious code and to provide controls against such code for which signatures do not yet exist or for which existing signatures may not be effective. Malicious code for which active signatures do not yet exist or may be ineffective includes polymorphic malicious code (i.e., code that changes signatures when it replicates). Nonsignature-based mechanisms also include reputation-based technologies. In addition to the above technologies, pervasive configuration management, comprehensive software integrity controls, and anti-exploitation software may be effective in preventing the execution of unauthorized code. Malicious code may be present in commercial off-the-shelf software as well as custom-built software and could include logic bombs, backdoors, and other types of attacks that could affect organizational mission and business functions.

In situations where malicious code cannot be detected by detection methods or technologies, organizations rely on other types of controls, including secure coding practices, configuration management and control, trusted procurement processes, and monitoring practices to ensure that software does not perform functions other than the functions intended. Organizations may determine that, in response to the detection of malicious code, different actions may be warranted. For example, organizations can define actions in response to malicious code detection during periodic scans, the detection of malicious downloads, or the detection of maliciousness when attempting to open or execute files.

**SL5 Supplemental Guidance:**

Adversarial content in AI model inputs (poisoned training data, jailbreak attempts, adversarial examples) is treated as malicious code. Standard malicious code protection mechanisms apply, including automated detection updates, alerting, and false positive handling.

## SI-3(10): Malicious Code Analysis

**NIST Control Text:**

a. Employ the following tools and techniques to analyze the characteristics and behavior of malicious code: [Assignment: organization-defined tools and techniques] ; and

b. Incorporate the results from malicious code analysis into organizational incident response and flaw remediation processes.

**NIST Discussion:**

The use of malicious code analysis tools provides organizations with a more in-depth understanding of adversary tradecraft (i.e., tactics, techniques, and procedures) and the functionality and purpose of specific instances of malicious code. Understanding the characteristics of malicious code facilitates effective organizational responses to current and future threats. Organizations can conduct malicious code analyses by employing reverse engineering techniques or by monitoring the behavior of executing code.

**SL5 Supplemental Guidance:**

The organization retains samples of detected adversarial content for analysis to understand attack mechanisms and improve detection capabilities. Analysis techniques vary based on attack type and available research, as effective analysis methods for adversarial AI content remain an active research area.

Analysis results feed continuous improvement: updating detection systems, refining thresholds, and identifying detection vulnerabilities. The organization tracks attack patterns and collects indicators of compromise (IOCs) from internal detections and external threat intelligence sources, distributing IOCs to detection developers, reviewers, and security operations.

The organization tracks which data was screened with which detector versions to enable re-scanning when detection capabilities improve significantly.

## SI-7(9): Verify Boot Process

**NIST Control Text:**

Verify the integrity of the boot process of the following system components: [Assignment: organization-defined system components].

**NIST Discussion:**

Ensuring the integrity of boot processes is critical to starting system components in known, trustworthy states. Integrity verification mechanisms provide a level of assurance that only trusted code is executed during boot processes.

**SL5 Supplemental Guidance:**

AI accelerators verify boot integrity using hardware-based mechanisms rooted in the hardware root-of-trust. Boot measurements are stored for attestation (IA-3), enabling remote verification that the accelerator booted with authorized firmware. Boot process verification for other system components is specified by applicable frameworks (CNSSI 1253 specifies boot integrity verification for all components that support it) and organizational policies.

## SI-7(10): Protection of Boot Firmware

**NIST Control Text:**

Implement the following mechanisms to protect the integrity of boot firmware in [Assignment: organization-defined system components]: [Assignment: organization-defined mechanisms].

**NIST Discussion:**

Unauthorized modifications to boot firmware may indicate a sophisticated, targeted attack. These types of targeted attacks can result in a permanent denial of service or a persistent malicious code presence. These situations can occur if the firmware is corrupted or if the malicious code is embedded within the firmware. System components can protect the integrity of boot firmware in organizational systems by verifying the integrity and authenticity of all updates to the firmware prior to applying changes to the system component and preventing unauthorized processes from modifying the boot firmware.

**Parameter Values:**

- Assignment (system components): AI accelerators within Weight Enclaves
- Assignment (mechanisms): Cryptographic verification using manufacturer-provisioned keys; hardware-protected firmware storage

**SL5 Supplemental Guidance:**

AI accelerators accept only manufacturer-signed firmware, verified using keys embedded in hardware during manufacturing. The accelerator rejects unsigned or incorrectly signed firmware even with full host system access.

Firmware protection is essential because firmware implements all other security mechanisms. If an attacker can modify firmware, they can disable memory isolation, attestation, and encrypted data paths.

## SI-7(15): Code Authentication

**NIST Control Text:**

Implement cryptographic mechanisms to authenticate the following software or firmware components prior to installation: [Assignment: organization-defined software or firmware components].

**NIST Discussion:**

Cryptographic authentication includes verifying that software or firmware components have been digitally signed using certificates recognized and approved by organizations. Code signing is an effective method to protect against malicious code. Organizations that employ cryptographic mechanisms also consider cryptographic key management solutions.

**Parameter Values:**

- Assignment (components): All code executing on AI accelerators within Weight Enclaves, including operator binaries

**SL5 Supplemental Guidance:**

AI accelerators verify that code is cryptographically signed before execution. This extends beyond firmware to operator binaries composing workload execution.

The organization ensures that approved workloads cannot be exploited by a compromised host to exfiltrate confidential information. Beyond verifying that individual operations do not leak data, this requires preventing composition attacks where small operations are sequenced to construct exfiltration channels.

Example approaches include task sequence verification, where firmware ensures operations execute in an approved sequence preventing a host from constructing malicious workflows, and restricting workloads to fused kernels that have been verified to not exfiltrate data individually or through composition.

Organizations implement continuous attestation verification with tamper-evident logging.

# Appendix A: Additional Open Questions

Organized by security stream:

## Machine Security

- Availability of accelerators meeting SL5 hardware security requirements in time for deployment
- Feasibility of workload integrity mechanisms (e.g., task sequence attestation, kernel fusion) for expected workloads such as training, inference, fine-tuning, and mechanistic interpretability

## Cryptographic Protection

- NSA Type 1 vs FIPS 140-3 Level 3 sufficiency for the stated threat model

## Physical Security

- TEMPEST zone classification for Red Zones
- Shared infrastructure isolation (power, cooling, fire suppression) between distributed Red Zones
- Required attenuation levels for shielded racks (full NSA 94-106 vs reduced spec)
- Level of physical isolation required for Weight Enclaves (rack, room, or building level separation)
- ICD 705 private accreditation: This standard specifies self-certification by CSO or Site Security Manager. Whether private self-certification provides meaningful security assurance compared to government accreditation, and what this implies for actual security posture

## Network Security

- Specific bandwidth thresholds for exfiltration prevention (calibrated to model size and threat model)
- Redundancy architectures for fail-closed encryptors

## Personnel Security

- Screening equivalency standards for non-US citizens and international offices
- Monitoring feasibility in high-restriction states (California CPRA/ICRAA, Illinois BIPA)

## Supply Chain Security

- Testing/certification approaches for adversarial content detection system robustness
- Processing state tracking for new modality onboarding